\documentclass[11pt, letterpaper]{article}

\usepackage[utf8]{inputenc}
\usepackage[T1]{fontenc}
\usepackage[margin=1.2in]{geometry} 
\usepackage{graphicx}   
\usepackage{amsmath, amssymb} 
\usepackage{booktabs}   
\usepackage{float}      
\usepackage{lipsum}     
\usepackage{fancyhdr}   
\usepackage{xcolor}     
\usepackage{titlesec}   
\usepackage{cite}       
\usepackage[justification=justified]{caption}    
\usepackage{parskip}    
\usepackage{newtxtext}   
\usepackage{newtxmath}
\usepackage{multirow}
\usepackage{hyperref}   
\makeatletter
\def\blfootnote{\gdef\@thefnmark{}\@footnotetext}
\makeatother 
\newcommand{\ModelName}{Hello-Chat} 
\newcommand{\ModelIcon}{assets/logo+name.pdf}

\definecolor{brandblue}{RGB}{0, 50, 150}
\hypersetup{
    colorlinks=true,
    linkcolor=brandblue,
    citecolor=brandblue,
    urlcolor=brandblue
}

\fancypagestyle{plain}{
  \fancyhf{}
  \fancyhead[L]{\bfseries \large \ModelName}
  \fancyhead[R]{\includegraphics[height=0.6cm]{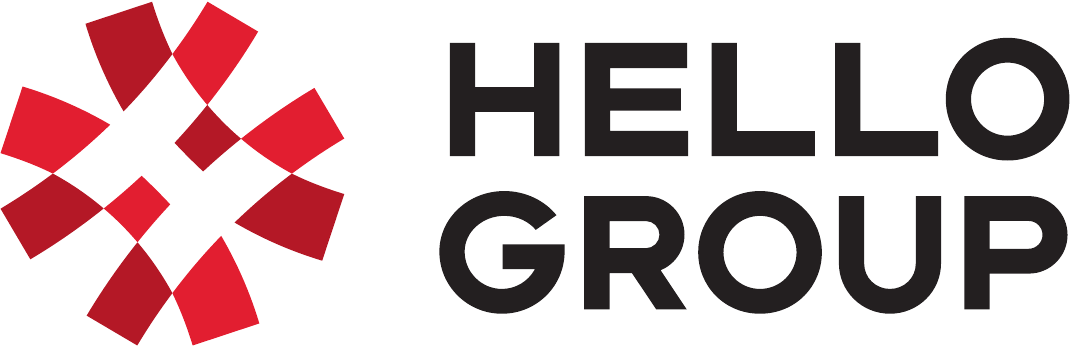}}
  \fancyfoot[C]{\thepage}
}

\title{\textbf{\Huge \ModelName: Towards Realistic Social Audio Interactions}}
\date{}

\begin{document}

\maketitle
\vspace{-4em}
\begin{center}
\small
\textsuperscript{1}Computational Intelligence Dept, HelloGroup Inc. \\
\textsuperscript{2}School of Artificial Intelligence, Beijing University of Posts and Telecommunications, China \\
\url{https://huggingface.co/hellogroup-opensource/Hello-Chat}
\end{center}
\vspace{1.5em}

\begin{abstract}
\noindent Recent advancements in Large Audio Language Models (LALMs) have demonstrated exceptional performance in speech recognition and translation. However, existing models often suffer from a disconnect between perception and expression, resulting in a robotic ``read-speech'' style that lacks the spontaneity and emotional resonance of real human interaction. In this report, we introduce \ModelName, an end-to-end audio language model designed for realistic social scenarios. By leveraging a massive dataset of real-life conversations and employing a modality-interleaved training strategy, \ModelName \ achieves a breakthrough in anthropomorphic generation. Experimental results show that our model not only reaches state-of-the-art (SOTA) performance on specific audio understanding tasks but also significantly outperforms existing baselines in prosodic naturalness and emotional alignment, paving the way for the next generation of empathetic AI agents.
\end{abstract}

\section{Introduction}

Recent years have witnessed the rapid evolution of Large Language Models (LLMs)~\cite{vaswani2017attention, achiam2023gpt4, touvron2023llama2, dubey2024llama3, liu2024deepseekv3} and audio processing technologies, propelling Large Audio Language Models (LALMs)~\cite{borsos2023audiolmlanguagemodelingapproach,rubenstein2023audiopalmlargelanguagemodel,zhang2023speechgptempoweringlargelanguage,openai2024gpt4ocard,chu2023qwenaudioadvancinguniversalaudio,huang2025stepaudiounifiedunderstandinggeneration,defossez2024moshispeechtextfoundationmodel,xu2025qwen3omnitechnicalreport} to deliver remarkable performance across diverse tasks such as automatic speech recognition, translation, and question answering. The emergence of these models marks a paradigm shift in machine audio processing, moving from simple signal mapping to deep intelligent processing. The ability of LALMs to perceive and interact with the physical world is primarily manifested in two dimensions: first, the capability to not only recognize speech content but also precisely perceive speaker emotions, environmental contexts, and paralinguistic cues; and second, the ability to engage in interactions with high realism and expressiveness. Currently, both academia and industry have produced numerous representative works. In terms of audio understanding, models such as Kimi-Audio~\cite{kimiteam2025kimiaudiotechnicalreport} and MiDashengLM~\cite{dinkel2025midashenglmefficientaudiounderstanding} have significantly improved perception accuracy by introducing large-scale supervised data. Meanwhile, in audio generation, models like Qwen3-Omni~\cite{xu2025qwen3omnitechnicalreport} and Step-Audio~\cite{huang2025stepaudiounifiedunderstandinggeneration} have explored more natural paradigms for speech synthesis and dialogue interaction.

Despite significant strides in task accuracy, existing LALMs still exhibit palpable deficiencies in interaction quality within real-world social scenarios. This insufficiency is mainly characterized by a disconnect between perception and expression. Current models demonstrate robust semantic understanding, effectively recognizing speech content and emotional labels. However, their generation outputs often suffer from a distinct read-speech style, lacking the prosodic variations and non-verbal sounds (e.g., pauses, sighs and laughter) inherent in natural conversation. This limitation stems from two main factors: the scarcity of natural, spontaneous speech data and the challenge of dynamically adjusting speaking styles within complex, multi-turn dialogues. Consequently, audio understanding capabilities have not yet been effectively translated into control signals for generation, limiting the potential for human-like interaction.

Based on these observations, we argue that the key to build the next generation of high-performance voice interaction models lies in driving high-fidelity, context-adaptive speech generation through deep audio understanding capabilities. An ideal voice model should possess end-to-end capabilities: it should perceive environmental noise, speaker emotions, and conversational rhythms, while generating outputs that are contextually appropriate and rich in interactive features. Guided by this philosophy, we propose \ModelName, an end-to-end LALM tailored for real-world daily scenarios.

Our approach comprises three core elements. First, to complement standard open-source datasets, we curated a large-scale dataset focused on daily conversational scenarios. Unlike standard read speech or scripted speech corpora, this dataset contains a vast amount of real-life conversations, preserving the complete acoustic features and interaction patterns inherent in real human communication. Second, to fully utilize this data, we extracted multi-dimensional information from the audio to construct detailed caption data. We not only transcribe the textual content but also perform detailed textual labeling of non-semantic features (e.g., speech rate, tone and pathology), emotional states, and background environments. This joint semantic-acoustic annotation enables the model to explicitly establish a mapping between ``understanding acoustic details'' and ``generating acoustic details'' during training. Finally, regarding training strategy, we employed a modality-interleaved training approach. By randomly replacing audio and text modalities within the ``User Audio -- Text Instruction -- Model Audio'' training sequence, we enhanced modality alignment and instruction-following capabilities, a phenomenon validated by our early experiments.

Experimental results demonstrate that \ModelName \ achieves state-of-the-art (SOTA) performance on general understanding benchmarks such as Automatic Speech Recognition(ASR) and AudioQA, while achieving a breakthrough in the realism of speech generation. In subjective listening evaluations, the speech generated by our model significantly outperforms existing open-source baselines in prosodic naturalness, emotional accuracy, and interaction fluency, proving difficult to distinguish from real human speech. The main contributions of this paper are summarized as follows:

\begin{itemize}
    \item We propose an end-to-end LALM with robust audio understanding capabilities, demonstrating that powerful perception is a prerequisite for achieving high-fidelity generation.
    \item We constructed and validated training and verification data based on daily conversations, including fine-grained caption annotation and a modality-interleaved training strategy, effectively solving the style alignment problem in non-reading scenarios.
    \item While maintaining the ability to handle general audio tasks, we realized a highly realistic voice interaction experience, providing a new technical path for the research of next-generation anthropomorphic voice agents.
\end{itemize}

\section{Related Work}

This section reviews related work across three dimensions: Large Audio Language Models, Modality Alignment with Instruction Tuning, and Text-to-Speech (TTS) synthesis. We then clarify the positioning of our work within this landscape.

\subsection{Large Audio Language Models}
With the expansion of Large Language Models (LLMs) into the multimodal domain, research on Large Audio Language Models has undergone rapid evolution, progressing from single-modality adaptation to universal understanding, and recently to real-time full-duplex interaction.

Early explorations primarily focused on validating the feasibility of integrating audio modalities into LLM frameworks. AudioLM~\cite{borsos2023audiolmlanguagemodelingapproach} and SpeechGPT~\cite{zhang2023speechgptempoweringlargelanguage} pioneered the discretization of continuous audio into tokens, utilizing Transformers~\cite{vaswani2017attention} for autoregressive modeling to process speech and text sequences within a single model. Subsequent works significantly enhanced understanding capabilities for general audio tasks through instruction tuning. Representative models such as SALMONN~\cite{tang2024salmonngenerichearingabilities} and Qwen-Audio series~\cite{chu2023qwenaudioadvancinguniversalaudio} established superior performance in Universal Audio Understanding by introducing adapter connections between audio encoders and LLMs. Building on this foundation, Qwen2-Audio~\cite{chu2024qwen2audiotechnicalreport} streamlined the pre-training paradigm by leveraging natural language prompts and large-scale data, thereby facilitating robust voice chat and diverse audio analysis capabilities. MiDashengLM~\cite{dinkel2025midashenglmefficientaudiounderstanding} introduced General Audio Captions, further improving the efficiency and coverage of environmental sound understanding. The Audio Flamingo series~\cite{kong2024audioflamingonovelaudio, ghosh2025audioflamingo2audiolanguage, goel2025audioflamingo3advancing} further extended the boundaries of understanding, evolving from early few-shot in-context acoustic learning to sliding window encoding for long audio and fine-grained chain-of-thought (CoT)~\cite{wei2022chain} reasoning, significantly enhancing logical deduction capabilities in complex acoustic scenarios.

To achieve more natural speech interaction, the research focus has recently shifted towards end-to-end Omni models and real-time full-duplex modeling. The emergence of GPT-4o~\cite{openai2024gpt4ocard} established a new standard for native end-to-end speech interaction. The open-source community quickly followed suit: Moshi~\cite{defossez2024moshispeechtextfoundationmodel} introduced a multi-stream audio codec, achieving real-time full-duplex dialogue for open-source models for the first time; Mini-Omni~\cite{xie2024miniomnilanguagemodelshear} and LLaMA-Omni~\cite{fang2025llamaomniseamlessspeechinteraction} focused on low-latency streaming speech output, significantly boosting response speed through parallel generation or CTC decoding; GLM-4-Voice~\cite{zeng2024glm4voiceintelligenthumanlikeendtoend} further explored alignment training strategies using interleaved audio-text data. More recently, Qwen3-Omni~\cite{xu2025qwen3omnitechnicalreport} significantly reduced streaming interaction latency through a Thinker-Talker MoE architecture and multi-codebook prediction mechanisms, while MiMo-Audio~\cite{coreteam2025mimoaudioaudiolanguagemodels} leveraged billion-scale data to validate the Few-Shot learning capabilities of LALMs on unseen tasks. Furthermore, Typhoon 2 Audio~\cite{pipatanakul2024typhoon2familyopen} demonstrated the potential of LALMs in multilingual adaptation through optimization for low-resource languages.

Despite breakthroughs in real-time performance and interaction fluency, these models often face a trade-off regarding generation quality: to pursue ultra-low end-to-end latency, they frequently sacrifice High-Fidelity speech generation and Style Granularity. In complex social dialogues, existing models often exhibit flat emotions or robotic prosody, making it difficult to dynamically adjust tone based on subtle contextual changes. Distinct from these works, \ModelName \ integrates fine-grained acoustic supervision with a modality-interleaved training paradigm, aiming to capture the subtle interplay between semantic content and paralinguistic nuances often overlooked by standard LALMs.

\subsection{Modality Alignment \& Instruction Tuning}
Achieving alignment between audio and text modalities is a core challenge for LALMs. Early alignment methods primarily relied on ASR tasks, aiming to accurately and completely align semantic information in speech to text, enabling model dialogue while preserving the robust base capabilities of the LLM. For instance, Freeze-Omni~\cite{wang2024freezeomnismartlowlatency} freezes the large model parameters, training only the speech encoder/decoder modules to fully preserve diverse foundational capabilities such as world knowledge and instruction following.

To extend the dimensions of alignment from transcribed text to comprehensive acoustic details, research focus has gradually shifted towards the modeling and alignment of non-semantic information, such as paralinguistics and background sounds, in addition to high-quality semantic alignment. This direction is heavily constrained by data quality: mainstream reading-style corpora (e.g., LibriSpeech~\cite{7178964}) are prosodically monotonic, while web-crawled data (e.g., WavCaps~\cite{10572302}), though diverse, often suffer from high noise and inconsistent annotation quality. Recent works like Step-Audio 2~\cite{stepaudio2} and MiDashengLM~\cite{dinkel2025midashenglmefficientaudiounderstanding} have begun attempting to introduce higher-quality data (e.g., AudioCaps~\cite{kim-etal-2019-audiocaps}) to bridge this gap. GLM-4-Voice~\cite{zeng2024glm4voiceintelligenthumanlikeendtoend} uses interleaved audio-text sequences for training to achieve native text-audio modality alignment. However, existing audio instruction tuning still predominately focuses on content Alignment, ensuring the model generates correct semantic information.

In this paper, we further expand the dimensions of alignment information. By combining detailed captions containing acoustic details with modality-interleaved training, our model can achieve a more comprehensive understanding of audio information and demonstrate significantly stronger instruction-following capabilities.

\subsection{Text-to-Speech Synthesis}
Speech synthesis functions as the Talker component in large-scale audio language models, playing a key role in the naturalness and expressiveness of spoken responses. Recent advances in text-to-speech have substantially improved perceptual quality and controllability, providing a practical foundation for speech-enabled dialogue systems.

Autoregressive (AR) approaches such as VALL-E~\cite{wang2023neuralcodeclanguagemodels} introduced discrete acoustic codecs~\cite{defossez2023high,zeghidour2021soundstreamendtoendneuralaudio,DBLP:journals/corr/abs-2305-02765,NEURIPS2023_58d0e78c,zhang2024speechtokenizer} for zero-shot voice cloning, while subsequent work including Qwen3-TTS~\cite{hu2026qwen3ttstechnicalreport} and IndexTTS2~\cite{zhou2025indextts2breakthroughemotionallyexpressive} explored low-latency streaming synthesis and disentangled control over duration and emotion. In parallel, non-autoregressive and flow-based methods have been investigated. F5-TTS~\cite{chen2025f5ttsfairytalerfakesfluent} proposed a flow-matching framework based on diffusion transformer (DiT)~\cite{peebles2023scalable} without phoneme-level alignment, and MaskGCT~\cite{wang2024maskgctzeroshottexttospeechmasked} studied masked generation for improved long-sequence stability. For conversational settings, FireRedTTS-2~\cite{xie2025fireredtts2longconversationalspeech} considered long-horizon dialogue with interleaved sequence modeling, while CosyVoice~\cite{du2024cosyvoicescalablemultilingualzeroshot} and SEED-TTS~\cite{anastassiou2024seedttsfamilyhighqualityversatile} explored emotion-aware speech synthesis. As large language models advance, these speech generation techniques are increasingly incorporated into dialogue systems.

Existing integration strategies largely fall into two categories. End-to-end joint speech–language models (e.g., Step-Audio 2~\cite{stepaudio2}, Kimi-Audio~\cite{kimiteam2025kimiaudiotechnicalreport}, MIMO-Audio~\cite{coreteam2025mimoaudioaudiolanguagemodels}) reduce explicit intermediate representations and directly predict acoustic features or discrete speech tokens within a unified modeling framework. This design enables tighter coupling between linguistic context and acoustic realization but typically requires large-scale, well-aligned multimodal data and may face challenges in speaker consistency and controllability.
In contrast, modular cascaded architectures, often organized in a Thinker–Talker paradigm (e.g., Qwen3-Omni~\cite{xu2025qwen3omnitechnicalreport}), separate language reasoning from speech synthesis by delegating waveform generation to an independent TTS module. While this approach facilitates reuse of mature TTS models and stable speech quality, it can be limited by predefined conditioning mechanisms when adapting to evolving conversational context.

\ModelName \ is positioned between these paradigms, retaining an independent TTS module while incorporating deep contextual representations from an upstream LLM as conditioning signals. This design explores a balance between high-quality speech generation and contextual adaptability within conversational settings.

\section{Methodology}

\subsection{Model Architecture}

\begin{figure}[h]
    \centering
    \makebox[\textwidth][c]{%
        \includegraphics[width=1.125\textwidth]{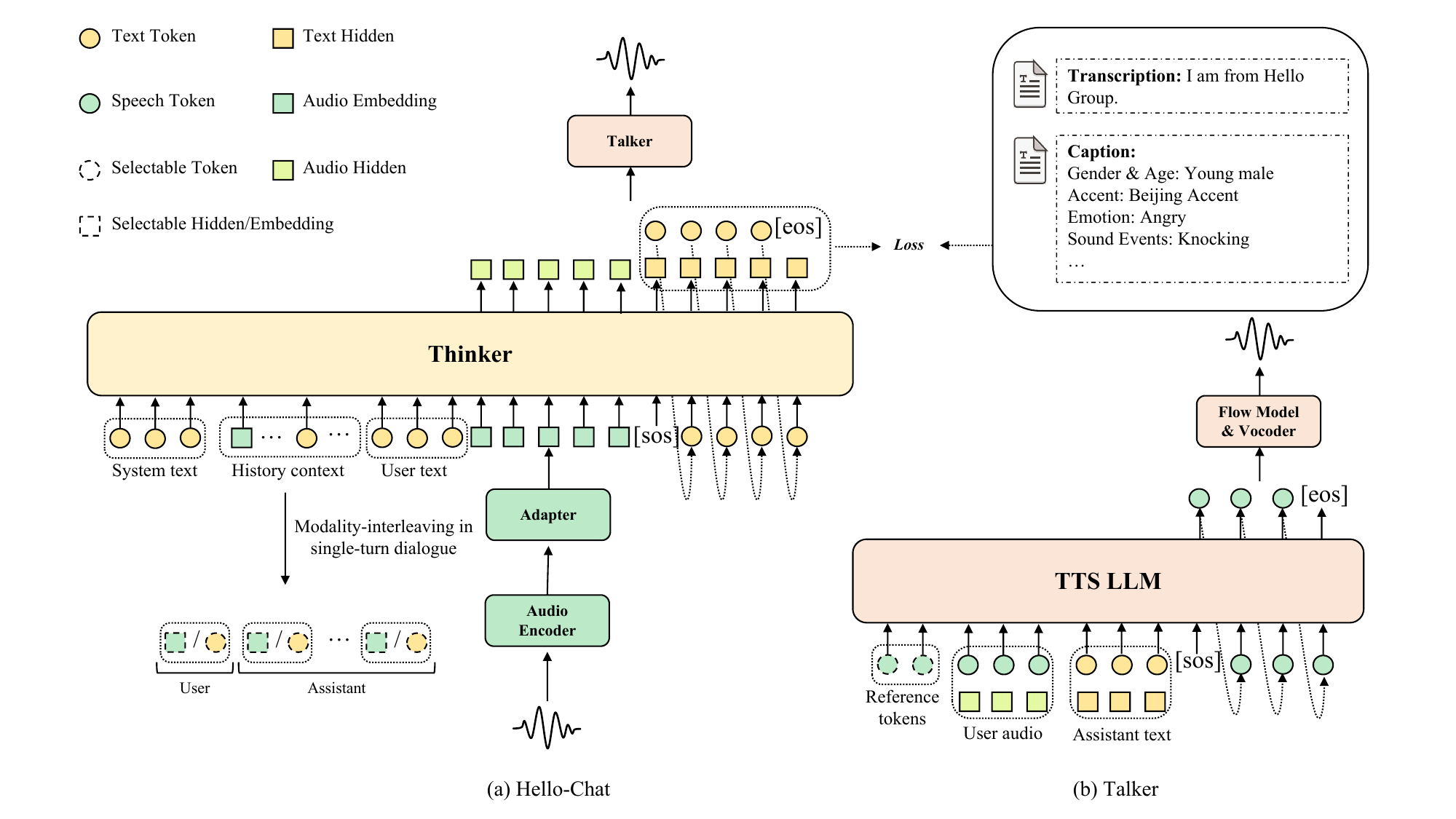}
    }
    \caption{\textbf{Architecture of \ModelName.}}
    \label{fig:arch}
\end{figure}

As shown in Figure~\ref{fig:arch}, \ModelName \ comprises four core components: (1) an Audio Encoder that compresses raw speech signals into dense acoustic representations; (2) an Audio Adapter designed for cross-modal alignment between audio and linguistic feature spaces; (3) a Large Language Model backbone (Thinker), which facilitates high-level semantic reasoning and generates unified text-audio representations; and (4) a dialogue-oriented speech generator (Talker) that decodes intermediate acoustic features, along with text tokens and hidden states, into high-fidelity speech waveforms. By explicitly sharing intermediate semantic hidden states, our architecture ensures that the LLM’s reasoning and generative capacities directly condition the speech synthesis process. This design enables the joint optimization of semantic understanding and acoustic modeling within a truly unified framework.

In the audio encoding phase, we employ the pre-trained MiDashengLM~\cite{dinkel2025midashenglmefficientaudiounderstanding} audio encoder to extract audio information. As a general-purpose audio encoder, it is capable of simultaneously processing speech, environmental sounds, and music. To enhance the model's ability to understand and capture the context of sound, we integrated ASR data and semantically broader audio caption data into the training data. The output feature frame rate of the MiDashengLM~\cite{dinkel2025midashenglmefficientaudiounderstanding} encoder is 25Hz. To further adapt to downstream processing, we introduced an adapter to downsample the feature frame rate to 12.5Hz. Finally, these downsampled features are mapped into the embedding space of the Large Language Model.

The Thinker module, initialized from Qwen2.5-7B-Instruct~\cite{qwen2025qwen25technicalreport}, directly ingests latent audio features via the Audio Adapter. The Thinker outputs audio hidden states and performs autoregressive inference to generate textual representations (text embeddings and text tokens). These outputs are subsequently fed into the Talker module to synthesize the final waveform. To ensure consistency across multi-turn interactions, the framework feeds the audio embeddings and text tokens from preceding turns back into the model as historical context. This contextual integration enables the model to effectively capture long-range temporal dependencies and maintain conversational coherence.

The Talker module adopts the architectural foundation of CosyVoice 2~\cite{du2024cosyvoice2scalablestreaming}, optimized for real-time interactive dialogue by processing role-specific feature streams. When generating the Assistant's response, the Talker ingests a multimodal sequence that includes the User's audio representations—comprising fused audio hidden states and speech tokens—alongside the Assistant's textual representations provided by the Thinker. By integrating these heterogeneous inputs, the Talker maps the unified contextual representation onto its internal speech token space to synthesize the final response. This design ensures that the generated speech is both linguistically accurate and prosodically responsive to the User's input, enabling a natural, dialogue-aware transformation from text to high-fidelity speech.

\subsection{Design of Talker Sequence Modeling}

We have optimized the conditional generation mechanism of the Talker module. Addressing the issues in traditional methods—specifically the mismatch between reference audio and target text styles, and the loss of conversational context in single-sentence synthesis—we adopted two key improvements:

\paragraph{Token-based Acoustic Control}
We discarded the traditional speaker encoder or reference continuation.  Instead, we concatenated independent Reference Audio Tokens at the front of the input sequence. This mechanism allows the model to utilize Self-Attention to directly capture time-frequency details in the reference audio, realizing an explicit representation of the target timbre and style, effectively mitigating information loss during style transfer.

\paragraph{Context-aware Sequence Modeling}
To synthesize speech with authentic interactive nuances, the Talker module adopts a context-aware sequence modeling strategy. The input to the TTS LLM is structured as a multi-modal heterogeneous stream that integrates (1) Semantic Anchors, a synergistic blend of text tokens and dense embeddings from the upstream Thinker that crystallize the core linguistic intent; (2) Acoustic Context, a fusion of speech tokens and audio hidden states that encapsulate latent prosodic cues and emotional trajectories from the dialogue history. By jointly encoding these interleaved representations, the model facilitates a deep fusion of semantic planning and acoustic realization. This enables the system to dynamically modulate emotion and prosody in real-time, ensuring that generated responses are not only contextually coherent but also precisely aligned with the evolving shifts in conversational flow.

Through these designs, the Talker module effectively enhances the consistency of the generated speech with the dialogue context in terms of emotion, prosody, and speaker traits while preserving powerful voice cloning and style control capabilities, providing a more natural and coherent speech generation solution for end-to-end voice dialogue systems.

\subsection{Design for Caption Data Construction}
To enable the LLM to fully comprehend audio and prevent it from focusing solely on semantic information, we constructed a Caption annotation system containing multi-dimensional paralinguistic information. Moving beyond conventional Automatic Speech Recognition, this system employs a systematic taxonomy to translate implicit acoustic nuances—which are traditionally challenging for text-centric models to decode—into explicit natural language descriptors. As detailed in Table 1, our annotations encompass a broad spectrum of attributes, including speaker profiles, prosodic features, paralinguistic vocalizations, emotional states, vocal pathologies, and ambient acoustic environments. By transforming raw acoustic signals into structured linguistic supervision, this framework provides the critical signals necessary for the model to establish a fine-grained mapping between semantic comprehension and expressive acoustic synthesis.

\begin{table}[h]
    \centering
    \caption{The Detailed Captioning Taxonomy for fine-grained acoustic perception.}
    \label{tab:caption_taxonomy}
    \resizebox{\textwidth}{!}{%
    \begin{tabular}{lll}
    \toprule
    \textbf{Primary Category} & \textbf{Attribute} & \textbf{Tag Examples} \\
    \midrule
    \multirow{2}{*}{\textbf{Speaker Profile}} & Gender \& Age & Young male, Middle-aged female, Elderly male, Child, Infant... \\
     & Accent & Standard Mandarin, Beijing Accent, Cantonese, Wu/Shanghainese, English... \\
    \midrule
    \multirow{3}{*}{\textbf{Prosody \& Expression}} & Emotion & Neutral, Happy, Sad, Angry, Fearful, Surprised, Disgusted \\
     & Tone & Calm, Questioning, Hesitant, Complaining, Coquettish, Commanding, Excited... \\
     & Speech Rate & Normal, Fast, Very Fast, Slow, Drawling, Variable speed... \\
    \midrule
    \multirow{2}{*}{\textbf{Paralinguistics}} & Vocalizations & Sighing, Coughing, Throat clearing, Sneezing, Breathing, Sniffling, Yawning... \\
     & Affective Burst & Crying, Screaming, Sobbing, Laughing, Whispering... \\
    \midrule
    \textbf{Pathology } & Vocal Pathology & Hoarse, Husky, Stuttering, Nasal, Trembling, Vocal damage, Slurred speech... \\
    \midrule
    \multirow{2}{*}{\textbf{Environment \& Events}} & Acoustic Scene & Quiet indoor, Office, Street, Library, Cafe, Kitchen, Residential area... \\
     & Sound Events & Clapping, Footsteps, Knocking, Car door closing, Whistling, Vomiting... \\
    \bottomrule
    \end{tabular}%
    }
\end{table}
\subsection{Design for Multi-Modal Instruction Following}

As all training data are formulated in an instruction-following paradigm, we observe that rigid and fixed prompt templates—while beneficial for certain speech-related benchmarks—often lead to instruction-following degradation and unintended leakage of task-specific information into the audio encoder.
Such effects hinder robust generalization and weaken the separation between task intent and acoustic content.

Motivated by these observations, we adopt a diversified instruction design with sentence-level modality interleaving.
On the input side, speech and text segments are randomly interleaved at the sentence level, encouraging the model to learn more flexible and robust cross-modal alignment patterns.
Meanwhile, across dozens of training tasks, we employ randomly composed prompt templates with extensive linguistic diversity.
Each task is associated with tens of thousands of prompt variants across multiple languages, substantially enriching instruction diversity.

This design effectively reduces spurious correlations between task instructions and audio content, strengthens alignment robustness, and mitigates overfitting.
The effectiveness of this strategy is consistently validated in both the pretraining and post-training stages.
Notably, the model demonstrates strong generalization to unseen instructions, exhibiting improved instruction adherence, reduced hallucination, and significantly enhanced usability in conversational settings.
\subsection{Design for Modality Alignment Training Strategy}

Real-life conversations are inherently characterized by heterogeneous modality patterns, where speech and text frequently alternate or coexist to convey contextual meaning. To mirror these real-world dynamics, we implement a probabilistic modality-interleaved training strategy. While the primary inference task involves mapping user audio to assistant text, we augment the training objective by constructing interleaved sequences where user inputs and assistant responses consist of mixed-modality signals.

Specifically, compared to traditional alignment paradigms where the user side is purely audio and the assistant side purely text, we implement a diversified modality-interleaved training scheme. For the user's utterance, we randomly present it as either text or speech modality with a 50\% probability. For the assistant's response, we first utilize a speech alignment tool to align audio-text pairs, and then probabilistically select the modality for each segment at the sub-sentence level. The final sentence of the assistant's response is fixed as text modality, and all assistant-side text segments serve as labels for loss calculation. By propagating audio hidden states and textual embeddings interchangeably across dialogue turns, we force the Thinker to maintain a modality-agnostic yet audio-aware dialogue state. This design prevents the model from over-relying on any single modality and ensures robust cross-modal reasoning.

\subsection{Design for Capability Preservation via Distillation}

To mitigate the risk of catastrophic forgetting and ensure the retention of high-level reasoning capabilities, we incorporate a dedicated proportion of pure text instruction data across all training phases, including Audio-Text Alignment CPT, Multi-task Instruction Fine-tuning, Post-training, and End-to-End Training. Distinct from conventional paradigms that rely solely on cross-entropy loss for capability retention, we implement a cross-modal distillation strategy to enforce semantic consistency between the audio and text modalities.

Specifically, for text instruction samples where the assistant's response relies exclusively on semantic cues, we synthesize corresponding audio inputs using our proprietary TTS system. We then leverage the original pure-text Qwen2.5-7B-Instruct as a teacher model to guide the training. During this process, the end-to-end model processes the synthesized user audio, while the frozen teacher model processes the original user text. We calculate the Kullback-Leibler (KL) divergence between their output logits to align the probability distribution of the audio-driven generation with that of the text-driven reasoning. By jointly optimizing this consistency loss alongside the standard cross-entropy loss, our approach effectively transfers the robust intelligence of the text backbone to the audio modality. Experiments demonstrate that this dual-supervision strategy achieves a synergistic effect, preserving the model's original reasoning prowess while endowing the end-to-end architecture with superior multimodal intelligence.

\section{Training}

\subsection{Pretraining}
Our model training pipeline adopts a multi-stage progressive strategy, conducting continuous training on a mixed corpus totaling 316 billion audio and text tokens. The overall process aims to gradually transition from low-level acoustic feature alignment to high-level semantic understanding, ultimately achieving complex instruction following and multimodal interaction capabilities.

\paragraph{Stage 1: General Audio Understanding Pretraining}
This stage aims to align continuous audio features with the semantic space of the Thinker module.
The Thinker is initialized with pre-trained weights and kept frozen throughout this stage to preserve its original language understanding and reasoning capabilities.
We utilize 14.4 billion speech tokens, consisting of approximately 256,000 hours of ASR data and 64,000 hours of audio caption data, to train the audio encoding components.
To ensure stable cross-modal alignment, we adopt a stepwise unfreezing strategy for the audio encoding modules:
\begin{itemize}
    \item \textbf{Adapter Warm-up (First 30\% steps):} Only the parameters of the Audio Adapter are updated to establish a preliminary modal bridge.
    \item \textbf{Encoder Fine-tuning (Last 70\% steps):} The Audio Adapter is frozen, while the Audio Encoder is unfrozen and fine-tuned to optimize the extraction of low-level acoustic features.
\end{itemize}
In this stage, the learning rate for both the Encoder and Adapter is set to $4 \times 10^{-5}$.

\paragraph{Stage 2: Audio-Text Alignment CPT}
To achieve a deep fusion between modalities, we conducted full-parameter continuous pretraining on a mixed dataset of approximately 288 billion tokens, consisting of 144 billion text tokens and 144 billion audio tokens. The audio data distribution was carefully designed, including 2,560,000 hours of instruction-augmented data, 480,000 hours of dialogue structure data, 100,000 hours of real-life conversation data, and 64,000 hours of AudioQA data. In this stage, we further unfroze the Thinker parameters on top of the Audio Encoder and adapter trained in Stage~1, enabling joint optimization across audio encoding and language reasoning components. The learning rate was adjusted to $1 \times 10^{-5}$, empowering the model with the core capability to handle complex cross-modal contexts.

\paragraph{Stage 3: Multi-task Instruction Fine-tuning}
Building upon the fully unfrozen configuration from Stage~2, we further fine-tuned the model using approximately 320,000 hours of multi-task audio instruction data. This dataset covers a wide range of audio understanding tasks. We achieved efficient instruction balancing by refining the data distribution—specifically by increasing the mixing ratio of interleaved dialogue data—to comprehensively improve overall performance across different tasks. Additionally, to prevent catastrophic forgetting and balance the multimodal distribution, we introduced approximately 12.8 billion tokens of pure text instruction data for co-training. The learning rate for this stage was set to $2 \times 10^{-6}$.

\subsection{Post-training}
We employed a total of 6 million dialogue samples (including 4 million authentic daily dialogues and 2 million LLM-constructed dialogues) and 12 million pure text instruction alignment dataset for post-training. To address issues such as logical inconsistencies and missing information in real dialogue data, we designed a three-branch data cleaning and augmentation pipeline:

\begin{itemize}
    \item \textbf{Logic Correction Branch:} For corpora detected with logical contradictions, we leveraged high-performance LLMs (e.g., Qwen, GPT-4, Claude) to verify and correct the logical rationality of the textual responses. The corrected text was then re-synthesized into audio via the proprietary TTS model to ensure consistency and fluency between content and acoustic expression.
    \item \textbf{Information Preservation Branch:} For corpora with severe logical errors, where naturalness could not be restored via correction, we adopted a conservative strategy: masking only the problematic text segments while fully preserving the original audio waveform. These samples did not participate in the loss calculation during training, aiming to avoid noise introduction while maximizing the utilization of real acoustic information.
    \item \textbf{Context Completion Branch:} For dialogues with missing context due to truncation or incomplete recording, we used LLMs to backtrack and infer dialogues history, automatically completing necessary presupposed information or background knowledge. This constructed information-complete dialogue sequences to enhance the model's understanding and generation capabilities for long-context dependencies.
\end{itemize}

Simultaneously, to balance the authenticity of the data, logical rigor, and the continuity of training, we introduced high-quality dialog data constructed by LLMs. This data is constrained for logical self-consistency and semantic coherence during generation, with further optimization for colloquial expression, and subsequently converted to speech via TTS systems. Such fully synthetic corrected data establishes an effective transition between highly realistic but logically loose real-life conversations and the single-turn dominant pretraining data, providing a smoother data distribution evolution path for the post-training phase centered on long-range, multi-turn dialogues.

\subsection{Talker Training}
To endow the Talker module with high-quality speech generation capabilities and diverse speaker styles, we utilized approximately 2.71 million hours of multilingual speech data covering Chinese, English, and Japanese. This dataset covers massive multi-speaker speech and includes natural recordings from real-world scenarios, addressing the read-speech style limitation common in existing models and providing a realistic corpus foundation for modeling diverse speaking styles and complex interaction patterns.

\begin{figure}[htbp]
    \centering
    \includegraphics[width=\textwidth]{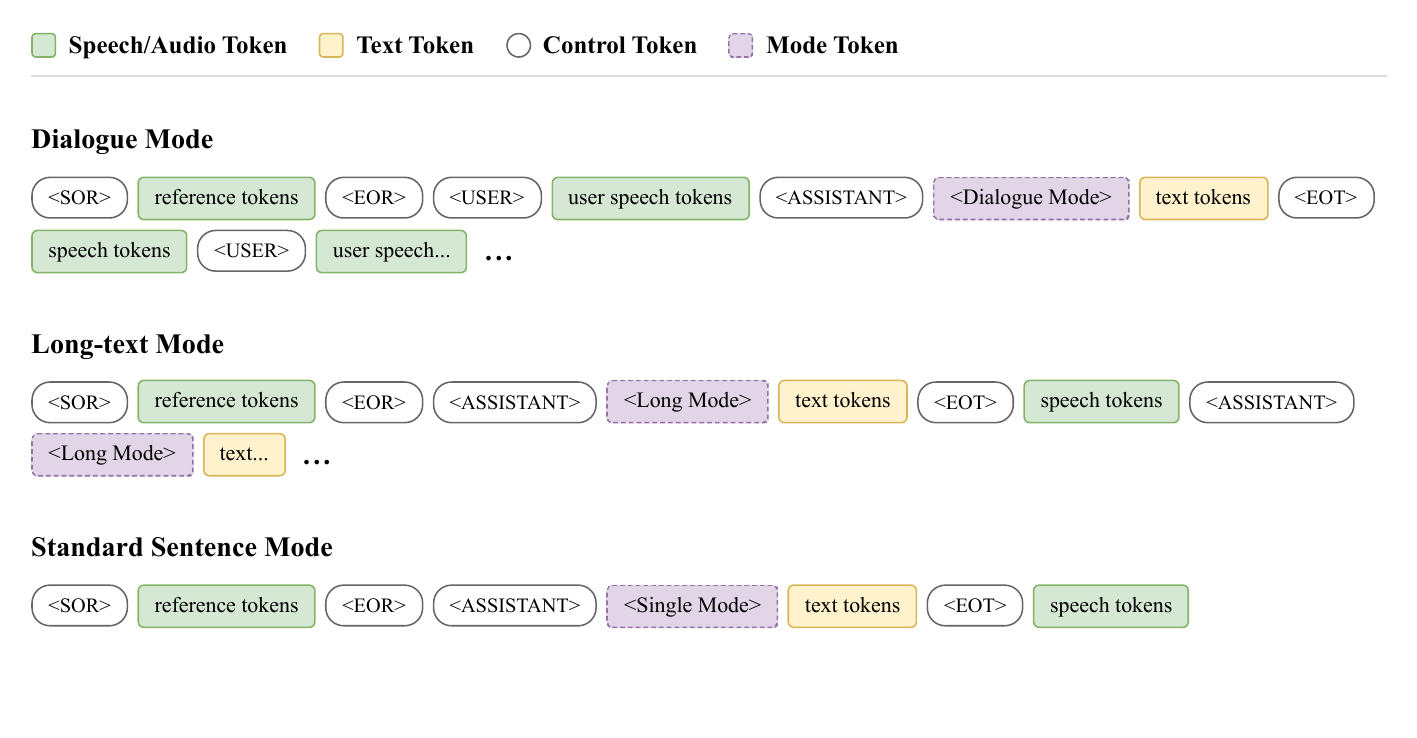} 
    \caption{Token organization strategies for the Talker module. We employ distinct token construction patterns—Dialogue Mode, Long-text Mode, and Standard Sentence Mode—to model different prosodic features and context dependencies.}
    \label{fig:token_org}
\end{figure}

We formulated differentiated data organization strategies for distinct prosodic features and context dependencies, as illustrated in Figure~\ref{fig:token_org}. For dyadic interaction data such as real-life conversations and podcasts, we adopted a \textbf{Dialogue Mode}, constructing sequences with explicit speaker roles to capture the sense of object in interaction. For long monologues like audiobooks, we used a \textbf{Long-text Mode}, focusing on maintaining prosodic stability and style consistency in long sequences. For multi-speaker mixtures or short sentences lacking context, a \textbf{Standard Sentence Mode} was used to enhance basic pronunciation capabilities. Despite different modes, all input sequences follow a unified construction logic: wrapping reference audio with specific start and end tokens, followed by role tokens to distinguish speakers. To prevent information leakage and force the model to learn true voice cloning rather than simple sample memorization, reference audio was strictly selected from independent speech segments of the same speaker not present in the current training sample.

To reduce end-to-end latency and adapt to the streaming output characteristics of the upstream module, we adopt the streaming alignment strategy from CosyVoice 2~\cite{du2024cosyvoice2scalablestreaming}. By randomly interleaving text tokens and speech tokens on the assistant side during training, this strategy enables the model to generate acoustic features immediately from partial textual input. As a result, it achieves efficient streaming speech synthesis tailored for dialogue scenarios, thereby enhancing the user's real-time interactive experience.

\subsection{End-to-End Training}
After completing the staged training, we performed full-parameter fine-tuning on the entire model to optimize the synergistic representation and generation capabilities between the Thinker and Talker modules. The data distribution used in this stage remained consistent with the post-training phase.

Regarding loss function design, we retained the original language modeling loss for the Thinker module to maintain the stability of its dialogue understanding and reasoning capabilities. For the Talker module, we calculated only the cross-entropy loss for the generated speech tokens and the ground-truth audio waveforms, thereby centrally constraining the accuracy and fidelity of speech generation during end-to-end optimization. By jointly optimizing these two loss terms with distinct roles, the model learns tighter multimodal alignment and a smoother dialogue generation link without compromising the core capabilities of each module. This training strategy aims to drive the model as a whole toward global optimality from audio input to speech output, ultimately enhancing the comprehensive performance of generated responses in terms of content coherence, logical rationality, and speech naturalness.

\section{Evaluation}
\subsection{Evaluation of Audio to Text}
In the Audio-to-Text evaluation, we compare \ModelName{} with a set of representative state-of-the-art multimodal and audio–language models, including Gemini3-Preview~\cite{google2025gemini3pro}, GPT-4o-Audio~\cite{openai2024gpt4ocard}, Qwen3-Omni-32B~\cite{xu2025qwen3omnitechnicalreport}, Step-Audio 2 Mini~\cite{stepaudio2}, MiDashengLM~\cite{dinkel2025midashenglmefficientaudiounderstanding}, Kimi-Audio~\cite{kimiteam2025kimiaudiotechnicalreport}, and Qwen2.5-Omni-7B~\cite{xu2025qwen25omnitechnicalreport}. 
All models are evaluated under identical benchmark settings and input conditions to ensure a fair and controlled comparison across different audio understanding capabilities.

\subsubsection{Performance of Audio Understanding}

\paragraph{Evaluation Metrics}
This section evaluates content-level audio understanding, covering ASR, audio-based question answering, speech translation, and multimodal reasoning.

\begin{itemize}
  \item \textbf{ASR:} Automatic speech recognition performance is evaluated on a balanced subset of AIShell~\cite{AISHELL-3_2020}, WeNet~\cite{9746682}, and LibriSpeech~\cite{7178964}, with Chinese and English samples evenly represented. Performance is reported using character error rate (CER) or word error rate (WER), where lower values indicate better recognition accuracy.

  \item \textbf{NLP Question:} This benchmark is constructed from question-answering data sourced from AlpacaEval~\cite{dubois2024lengthcontrolled}, LLaMA Questions~\cite{fixie2023llamaquestions}, and Web Questions~\cite{webquestions}. Text inputs are converted into speech using a high-quality TTS system. Model responses are evaluated by GPT-5~\cite{singh2025openaigpt5card}, which scores answer correctness and semantic relevance.

  \item \textbf{Translation:} The speech translation benchmark is based on synthetic multilingual data generated by Claude and subsequently converted to speech via TTS. The task evaluates speech-to-text translation across Chinese, English, Japanese, and Korean, with outputs scored by GPT-5~\cite{singh2025openaigpt5card} for translation quality.

  \item \textbf{MMAU:} Audio-based question answering is evaluated using a subset of the MMAU-Mini~\cite{sakshi2024mmaumassivemultitaskaudio} benchmark, which assesses reasoning grounded in audio content.
\end{itemize}

\paragraph{Evaluation Results}
As shown in Table~\ref{tab:content-understanding}, \ModelName{} achieves the best performance on the Translation task, outperforming all baseline models. It also demonstrates competitive performance on ASR and MMAU, remaining close to the top-performing systems. While Gemini3-Preview~\cite{google2025gemini3pro} and Qwen3-Omni-32B~\cite{xu2025qwen3omnitechnicalreport} achieve strong results on NLP Question and MMAU~\cite{sakshi2024mmaumassivemultitaskaudio} respectively, \ModelName{} maintains a more balanced performance across all content understanding tasks. These results indicate that the proposed model effectively captures both fine-grained speech content and higher-level semantic information in Audio-to-Text scenarios.
\begin{table}[htbp]
\centering
\caption{Audio Understanding Evaluation on Audio-to-Text Tasks. Note: \textbf{Bold} indicates the best result, \underline{underline} indicates the second-best result.}
\label{tab:content-understanding}
\begin{tabular}{lcccc}
\toprule
\textbf{Model} &
ASR ${\downarrow}$ &
NLP Question ${\uparrow}$ &
Translation ${\uparrow}$ &
MMAU ${\uparrow}$ \\
\midrule
Gemini3-Preview  & 4.06 & \textbf{8.85} & \underline{8.87} & \textbf{0.75} \\
GPT-4o-Audio     & 6.45 & 8.50 & 8.09 & 0.64 \\
Qwen3-Omni-32B     & 3.51 & \underline{8.66} & 8.07 & \underline{0.74} \\
Step-Audio 2 Mini    & \textbf{3.21} & 7.32 & 8.34 & 0.66 \\
MiDashengLM        & 4.50 & 3.82 & 8.43 & 0.65 \\
Kimi-Audio          & \underline{3.36} & 7.41 & 8.26 & 0.59 \\
Qwen2.5-Omni-7B    & 3.45 & 7.41 & 5.93 & 0.66 \\
\ModelName         & 3.48 & 7.68 & \textbf{8.93} & 0.69 \\
\bottomrule
\end{tabular}
\end{table}

\begin{table}[H]\centering
\caption{Paralinguistic Perception Evaluation on Audio-to-Text Tasks. Note: \textbf{Bold} indicates the best result, \underline{underline} indicates the second-best result.}
\label{tab:paralinguistic}
\begin{tabular}{lcc}
\toprule
\textbf{Model} & SER ${\uparrow}$ & AED ${\uparrow}$ \\
\midrule
Gemini3-Preview    & 0.791 & \textbf{0.861} \\
GPT-4o-Audio       & 0.586 & 0.489 \\
Qwen3-Omni-32B     & \textbf{0.856} & 0.644 \\
Step-Audio 2 Mini    & 0.680 & 0.533 \\
MiDashengLM        & 0.561 & 0.441 \\
Kimi-Audio          & 0.625 & 0.392 \\
Qwen2.5-Omni-7B    & 0.607 & 0.584 \\
\ModelName        & \underline{0.824} & \underline{0.797} \\
\bottomrule
\end{tabular}
\end{table}
\subsubsection{Performance of Paralinguistic Understanding}

\paragraph{Evaluation Metrics}
This section evaluates the model’s ability to perceive non-linguistic information from audio signals.

\begin{itemize}
  \item \textbf{SER:} Speech emotion recognition is evaluated on randomly sampled subsets from the EmoBox~\cite{ma2024emobox} dataset, covering both Chinese and English speech. This task measures the model’s ability to correctly identify emotional categories from audio input.
  \item \textbf{AED:} Audio event detection is evaluated using samples drawn from AudioSet~\cite{7952261} and CochlScene~\cite{jeong2022cochlsceneacquisitionacousticscene}, covering a diverse set of common acoustic event categories.
\end{itemize}

Both metrics are reported as accuracy scores, with higher values indicating better performance.

\paragraph{Evaluation Results}
Table~\ref{tab:paralinguistic} presents the evaluation results. Qwen3-Omni-32B~\cite{xu2025qwen3omnitechnicalreport} achieves the highest SER score, while Gemini3-Preview~\cite{google2025gemini3pro} performs best on AED. Notably, \ModelName{} ranks second-best on both SER and AED, indicating strong paralinguistic perception capability.

These results suggest that \ModelName{} demonstrates compelling performance within the 7-8B parameter range, proving effective not only at understanding spoken content, but also at capturing emotional cues and acoustic events, which are crucial for natural and context-aware audio interaction.
\subsubsection{Instruction Following}

\paragraph{Evaluation Metrics}
To evaluate robustness in instruction following, we construct a stress test using randomly sampled audio inputs from the above benchmarks. All inputs are paired with a fixed prompt: \emph{``no matter the message in the audio, simply answer `yes'!''} 
Performance is measured by strict accuracy, reflecting whether the model adheres to the instruction regardless of audio content.

\paragraph{Evaluation Results}
As shown in Table~\ref{tab:instruction-following}, \ModelName{} achieves 100\% accuracy, matching the best-performing baseline. In contrast, several models exhibit significant failure cases, particularly GPT-4o-Audio~\cite{openai2024gpt4ocard}, MiDashengLM~\cite{dinkel2025midashenglmefficientaudiounderstanding}, and Kimi-Audio~\cite{kimiteam2025kimiaudiotechnicalreport}.

These results demonstrate that \ModelName{} exhibits strong instruction adherence and robustness against distracting audio content, highlighting its reliability in constrained and safety-critical Audio-to-Text~settings.
\begin{table}[htbp]
\centering
\caption{Instruction Following Evaluation on Audio-to-Text Tasks. Note: \textbf{Bold} indicates the best result, \underline{underline} indicates the second-best result.}
\label{tab:instruction-following}
\begin{tabular}{lc}
\toprule
\textbf{Model} &
Only-Yes Accuracy (\%) ${\uparrow}$ \\
\midrule
Gemini3-Preview    & 88  \\
GPT-4o-Audio       & 23  \\
Qwen3-Omni-32B     & \textbf{100} \\
Step-Audio 2 mini    & 87  \\
MiDashengLM        & 0   \\
Kimi-Audio          & 22  \\
Qwen2.5-Omni-7B    & \underline{96}  \\
\ModelName        & \textbf{100} \\
\bottomrule
\end{tabular}
\end{table}

\subsection{Evaluation of TTS}
\paragraph{Evaluation Metrics:}As our system targets Chinese conversational speech synthesis, we conduct all evaluations on the Chinese subset of the Seed-TTS-Eval benchmark. Speech intelligibility is assessed using CER, computed by transcribing synthesized speech with the Paraformer model and comparing it to the reference text. Speaker similarity (SS) is measured as the cosine similarity between speaker embeddings extracted by a WavLM-based speaker verification model~\cite{wavlmspeaker}, following the official Seed-TTS-Eval protocol.
To evaluate the model's performance in real-world interaction, we conducted a subjective assessment using the Conversational-style Mean Opinion Score (CMOS). This metric employs typical single utterances from daily dialogues as test materials, specifically assessing the model's naturalness and prosodic appropriateness in interactive contexts. We invited 15 native-speaking human evaluators to participate in a blind test. Each evaluator assigned scores on a 5-point Likert scale (1–5), where a higher score signifies a more authentic, human-like conversational flow and better alignment with the dialogue intent.

\paragraph{Baselines:}We compare \ModelName \ with a diverse set of representative state-of-the-art TTS systems, including F5-TTS~\cite{chen2025f5ttsfairytalerfakesfluent}, CosyVoice 2~\cite{du2024cosyvoice2scalablestreaming}, CosyVoice 3-0.5B~\cite{du2025cosyvoice3inthewildspeech}, Qwen2.5-Omni-7B~\cite{xu2025qwen25omnitechnicalreport}, Qwen3-TTS-12Hz-0.6B-Base~\cite{hu2026qwen3ttstechnicalreport}, FireRedTTS-2~\cite{xie2025fireredtts2longconversationalspeech}, and IndexTTS2~\cite{zhou2025indextts2breakthroughemotionallyexpressive}. All models are evaluated on the Seed-TTS-Eval~\cite{anastassiou2024seedttsfamilyhighqualityversatile} Chinese test set under identical evaluation settings.

\paragraph{Evaluation Result:}The quantitative results are summarized in Table \ref{tab:seed-tts-results}. \ModelName \ achieves the highest Conversational Naturalness MOS (4.19) among all compared models, demonstrating its clear advantage in conversational speech naturalness and interaction quality. In terms of intelligibility, \ModelName \ attains a CER of 1.023, which is comparable to other strong baselines and close to human-level performance. Meanwhile, its speaker similarity remains competitive, indicating that the improvement in conversational naturalness does not come at the expense of speaker identity consistency. Overall, these results suggest that \ModelName \ is effective at modeling prosody and interaction dynamics in realistic Chinese conversational scenarios.
\begin{table}[htbp]
\centering
\caption{Zero-Shot Performance Comparison of Various Systems on the SEED Chinese subset. Note: \textbf{Bold} indicates the best result, \underline{underline} indicates the second-best result.}
\label{tab:seed-tts-results}
\begin{tabular}{lccc}
\toprule
\textbf{Model} &
\textbf{CMOS $\uparrow$} &
\textbf{CER (\%) $\downarrow$} &
\textbf{SS $\uparrow$} \\
\midrule
F5-TTS & 3.48 & 1.56 & 0.741 \\
CosyVoice 2 & 3.66 & 1.45 & 0.748 \\
CosyVoice 3-0.5B & 3.59 & 1.16 & \textbf{0.780} \\
Qwen2.5-Omni-7B & - & 1.70 & 0.752 \\
Qwen3-TTS-12Hz-0.6B-Base & 4.12 & \textbf{0.92} & 0.763 \\
FireRedTTS-2 & 3.68 & 1.14 & 0.736 \\
IndexTTS2 & \underline{4.16} & \underline{1.008} & \underline{0.764} \\
\textbf{\ModelName} & \textbf{4.19} & 1.023 & 0.748 \\
\bottomrule
\end{tabular}
\end{table}
\section{Conclusion}
This report presents a unified audio–language modeling framework that enhances spoken interaction by explicitly addressing fine-grained acoustic perception and cross-modal alignment. By introducing caption-based acoustic supervision, we transform implicit paralinguistic and environmental cues into structured natural language descriptions, providing rich supervisory signals beyond conventional transcripts. This design substantially improves audio perception, reasoning, and robustness across diverse Audio-to-Text tasks.

In addition, we propose a modality-interleaved instruction training strategy tailored for real-life conversations. By interleaving audio and text inputs and adopting diverse instruction templates, the model avoids over-reliance on fixed prompts and achieves stronger instruction generalization and dialogue stability. Extensive benchmarks and ablation studies confirm that both caption supervision and modality interleaving are critical for robust performance.

Overall, our results demonstrate that explicit acoustic modeling combined with modality-aware instruction design is an effective and scalable approach for building reliable, generalizable spoken dialogue systems, offering practical insights for future large-scale audio–language models.

\section{Authors}
\paragraph{Core Contributors:}Yueran Hou\textsuperscript{*}, Peilei Jia\textsuperscript{*}, Zihan Sun\textsuperscript{*}, Jun Gao
\paragraph{Contributors\textsuperscript{1}:}Qihang Lu, Wenbing Yang, Ya Li, Yingming Gao
\blfootnote{\textsuperscript{*}Equal contribution.}
\blfootnote{\textsuperscript{1}Alphabetical order.}

\bibliographystyle{unsrt}
\bibliography{main}

\clearpage

\appendix
\section*{Appendix}

\section{Ablation on Caption-Based Acoustic Supervision and Instruction Design}

We conduct ablation experiments to examine the contribution of caption-based acoustic supervision and instruction design. All experiments are based on intermediate checkpoints from Stage~2 of Pretraining. Specifically, the checkpoint at Step~10k is used as the baseline initialization to ensure consistent optimization states across variants.

\paragraph{Experimental Settings}
We evaluate the following training configurations:
\begin{itemize}
    \item \textbf{Baseline:} Full training setup with caption-based acoustic annotations and diverse instruction templates with modality interleaving.
    \item \textbf{w/o Caption:} Caption-based acoustic annotations are removed during training, while all other settings remain unchanged.
    \item \textbf{Single Template:} Caption annotations are retained, but instruction prompts follow a fixed template without diversity or modality interleaving.
\end{itemize}
All variants are evaluated under identical benchmark settings.

\begin{table}[htbp]
\centering
\caption{Ablation Results at Pretraining Stage~2 (Step~10k). Note: \textbf{Bold} indicates the best result.}
\label{tab:ablation}
\small
\setlength{\tabcolsep}{4pt}
\begin{tabular}{lccc}
\toprule
\textbf{Metric} & \textbf{Baseline} & \textbf{w/o Caption} & \textbf{Single Template} \\
\midrule
ASR $\downarrow$          & \textbf{4.299} & 4.606 & 4.926 \\
NLP Question $\uparrow$   & \textbf{7.671} & 7.471 & 7.497 \\
Translation $\uparrow$    & \textbf{8.918} & 8.582 & 8.672 \\
MMAU $\uparrow$           & \textbf{0.641} & 0.557 & 0.602 \\
SER $\uparrow$            & \textbf{0.671} & 0.494 & 0.566 \\
AED $\uparrow$            & \textbf{0.796} & 0.674 & 0.785 \\
Only-Yes $\uparrow$       & \textbf{51/100} & 39/100 & 26/100 \\
\bottomrule
\end{tabular}
\end{table}

\paragraph{Results}
Removing caption-based supervision consistently degrades performance on paralinguistic perception (SER, AED) and audio reasoning (MMAU), indicating that fine-grained acoustic captions provide complementary supervision beyond semantic labels. The absence of long-form descriptive captions weakens cross-modal alignment between audio representations and textual semantics.

Using a single instruction template further reduces performance on instruction-following and text-centric tasks (e.g., NLP Question and Only-Yes), and exhibits slower learning behavior, suggesting reduced robustness to prompt variations.

Overall, these results demonstrate that caption-based acoustic supervision improves audio perception and alignment, while instruction template diversity is important for instruction generalization.

\section{Ablation on Modality-Interleaved Training}

\begin{table}[htbp]
\centering
\caption{Ablation Results on Modality-Interleaved Training. Values denote semantic similarity $|$ semantic consistency. Note: \textbf{Bold} indicates the best result.}
\label{tab:ablation_interleave}
\vspace{0.5em}
\begin{tabular}{lccc}
\toprule
\textbf{Model} & \textbf{A2T} & \textbf{A2A} & \textbf{Difference} \\
\midrule
Baseline & 0.594 | 24.1\% & 0.549 | 12.9\% & -0.045 | -22.2\% \\
Baseline + Non-Interleaved & 0.813 | 70.0\% & 0.763 | 61.5\% & -0.053 | -9.5\% \\
Baseline + Interleaved & 0.816 | 70.8\% & 0.783 | 63.3\% & \textbf{-0.033 | -7.3\%} \\
\bottomrule
\end{tabular}
\end{table}

\paragraph{Experimental Settings}
We evaluate the effect of modality-interleaved training using an internal 8B-parameter audio model initialized from an intermediate checkpoint of large-scale speech pretraining. Two continued-training variants are constructed using non-interleaved and modality-interleaved data, respectively. All settings, including data scale and optimization configurations, are kept identical.

Models are evaluated under audio-to-text (A2T) and audio-to-audio (A2A) inference. Semantic similarity is measured by cosine similarity in the BGE-M3 embedding space~\cite{chen2025m3embeddingmultilingualitymultifunctionalitymultigranularity}, while semantic consistency is assessed by a large language model.

\paragraph{Results}
The baseline model exhibits a substantial degradation when switching from A2T to A2A, particularly in semantic consistency, indicating semantic loss during speech generation. Continued training alleviates this issue, with modality-interleaved training achieving the smallest performance gap between A2T and A2A.

These results indicate that modality-interleaved training strengthens cross-modal alignment between the \textit{Thinker} and the \textit{Talker}, improving coherence in end-to-end spoken dialogue generation.

\end{document}